\documentclass[conference]{IEEEtran}  

\IEEEoverridecommandlockouts
\usepackage[left=0.63in,right=0.63in,top=0.75in,bottom=1.1in]{geometry}
\def\BibTeX{{\rm B\kern-.05em{\sc i\kern-.025em b}\kern-.08em
    T\kern-.1667em\lower.7ex\hbox{E}\kern-.125emX}}
\usepackage[utf8]{inputenc}
\usepackage[backend=biber, style=ieee]{biblatex}
\usepackage{amsmath,amsfonts, amsthm}
\usepackage{algorithmic}
\usepackage{array}
\usepackage[caption=true,font=footnotesize,labelfont=sf,textfont=sf]{subfig}
\usepackage{tcolorbox}
\usepackage{color}
\usepackage{enumitem}
\usepackage{url}
\usepackage{graphicx}
\usepackage[hidelinks]{hyperref}
\usepackage{wrapfig}
\theoremstyle{definition}

\usepackage{orcidlink}
\usepackage[normalem]{ulem}
\usepackage{longtable, booktabs, etoolbox}
\usepackage{multirow}
\usepackage[noabbrev,nameinlink]{cleveref}
\bibliography{bibliography}  

\renewcommand{\itemautorefname}{\@gobble}
\usepackage{xspace}

\begin{document}
\title{Embedded Software Development with \\ Digital Twins: Specific Requirements for Small and Medium-Sized Enterprises}
%
%
\author{
\IEEEauthorblockN{Alexander Barbie\IEEEauthorrefmark{1}\IEEEauthorrefmark{2} and Wilhelm Hasselbring\IEEEauthorrefmark{2}} \\ 
\IEEEauthorblockA{\IEEEauthorrefmark{1}GEOMAR Helmholtz Centre for Ocean Research Kiel (Germany)}
\IEEEauthorblockA{\IEEEauthorrefmark{2}Software Engineering Group, Christian-Albrechts-University, Kiel (Germany)}
}
\IEEEaftertitletext{\vspace{-1.5\baselineskip}}
\IEEEtitleabstractindextext{%
\begin{abstract}
The transformation to Industry 4.0 changes the way embedded software systems are developed. Digital twins have the potential for cost-effective software development and maintenance strategies. With reduced costs and faster development cycles, small and medium-sized enterprises (SME) have the chance to grow with new smart products. We interviewed SMEs about their current development processes. In this paper, we present the first results of these interviews. First results show that real-time requirements prevent, to date, a Software-in-the-Loop development approach, due to a lack of proper tooling. Security/safety concerns, and the accessibility of hardware are the main impediments. Only temporary access to the hardware leads to Software-in-the-Loop development approaches based on simulations/emulators. Yet, this is not in all use cases possible. All interviewees see the potential of Software-in-the-Loop approaches and digital twins with regard to quality and customization. 
One reason it will take some effort to convince engineers, is the conservative nature of the embedded community, particularly in SMEs.
\end{abstract}

}

\maketitle

\IEEEdisplaynontitleabstractindextext
\section{Introduction}\label{sec:introduction}
The popularity of digital twins (DT) increased in the past several years. Both, for monitoring machines in production, and for the development process. With the development of DTs, advanced methods and architectures facilitate opportunities to change the competition in traditional markets, most notably in the automation and process technology. Introducing new or advanced processes in large companies with over time grown structures can be laborious. Especially, if various departments and decision makers are involved. Although these companies may have large development teams, they do not develop new products from scratch and mostly have to use enterprise-wide technology stacks.

In this environment, new opportunities for smaller and medium-sized enterprises (SME) arise. The definition by the European Commission for SMEs is as follows: A small enterprise, is a company with up to 49 employees and a yearly turnover of a maximum of 10 million Euro; A middle-sized enterprise, is a company with a maximum of 249 employees and a yearly turnover of a maximum of 50 million Euro \cite{SMEDefinition}. In Germany, these SMEs form the backbone of the economy. While Germany has global players like VW and Siemens, around $99\,$\% of all enterprises are SMEs and they have a big impact on the export of goods, in particular, the so called \emph{Hidden Champions}. Hidden champions are companies that have a huge revenue, are number one, two, or three in the global market, and yet have a low level of public awareness \cite{Simon}.
To emphasize the meaning of these German SMEs and Hidden Champions and their role for Industry 4.0 applications, we emphasize the numbers of the sensor and measurement technology manufacturers in 2017, published by the AMA Association for Sensors and Measurement in their yearly press review \cite{ama2017}. Around 2500 companies generated 35 billion Euro in annual sales. Furthermore, these enterprises hold a world market share of around $30$\%. Sensors are the smallest units of an embedded system and absolutely essential for the Industry 4.0 and especially, the German automotive sector. 

In the context of a Helmholtz Enterprise Field Study Fellowship program, we conducted semi-structured interviews with SMEs in, mostly, northern Germany. In this paper, we present the first results of these interviews. The interviews were conducted in German. We translated the questions and quotes to English for this paper.

\section{Related Work}\label{sec:relatedwork}
Researching the processes and the impact of new software architectures on the transformation to an Industry 4.0 is a broad field. Over the past years, many studies were published with regard to novel architectures and the utilization of DTs. Although the research scopes are different, they draw a broader picture of day to day challenges engineers face.

Product \emph{quality} is one of the key criteria for customer experience in the Industry 4.0. Achieving this in an environment with increasingly complex machines is a challenge. The fact that quality is perceived in the software industry as the single most relevant premise to survive \cite{50yearsse}, was also found in a survey among 2,000 decision makers about trends and challenges in software engineering. Yet, organizations struggle to achieve quality along with cost and efficiency \cite{SIsoftwarequality}. 
During the development of embedded (software) systems, at some point, thorough and reliable tests are necessary to verify and validate the whole system \cite{DTElectricalVehicles}. A common way to test the control algorithms of an embedded software system is Hardware-in-the-Loop (HIL) testing. An example for HIL testing at large scale is Airbus with creating iron birds of their aircrafts, containing the corresponding electrics, hydraulics and flight controls \cite{ironbird}. However, SMEs cannot afford such redundant hardware just for the purpose of testing software. Hence, test automation is among the most popular topics for testing embedded software \cite{studyembeddedtesting}. With increasing importance of automated testing, the research around its impact, more efficient workflows, and advanced tooling also increases.

\textcite{Zampetti2022} conducted an interview study with Continuous Integration / Continuous Delivery (CI/CD) practices for cyber physical system (CPS) with ten participants for semi-structured interviews combined with a survey, where 55 professional developers participated. Different to our work, the interviews also included large enterprises. An interesting result of the surveys is that more than half of the survey participants considered the presence of a development environment detached from the execution environment as a real impediment to set up a CI/CD process for CPSs \cite{Zampetti2022}. Developing software in an environment different from the production environment can lead to faulty implementations and hence, additional time and costs to fix the problems after deploying the embedded software system to the execution environment. The idea of DTs is to have a precise digital copy of the real physical asset. This means not only a digital model of the asset, but also the embedded software system. Our Digital Twin Prototype (DTP) (see \Cref{subsec:ResultRQ1} and \cite{MFI2020}) approach addresses this challenge for engineers. 

With regards to DTs, the definition of a DT differs between practice and academia. The major differences are visualized via the box models of \textcite{KritzingerDTBoxes}. Similar, \textcite{vanderValk2021} conducted a literature review to find archetypes of DTs used by practitioners in industry. In addition, they conducted an interview study. They conclude that their archetypes show an evolutionary process for DTs, from simple data integration to fully autonomous systems. An interesting result from the interviews is that interviewees argued that the practical agreement and regulatory aspects may hinder the realization of highly developed archetypes \cite{vanderValk2021}.

In another interview study, \textcite{Neto2020} elaborate drivers (e.g. flexibility requirements for new markets; competition pressures on costs, productivity, and quality), enablers (e.g. IoT, process standardization, and skills to manage the technologies), and barriers to the implementation. Some of the barriers result from a lack of maturity of the utilized technologies, but also with regard to people and competences. A possible barrier are missing qualifications and resistance to change, due to caution or fear of the unknown (see Fig. 2 in \cite{Neto2020}).

\section{Our Research Questions}
In previous research, we demonstrated how DTs can be applied to develop and monitor a network of ocean observation systems (OOS) \cite{fieldreport}. In the Helmholtz Future Project ARCHES (Autonomous Robotic Networks to Help Modern Societies) with a consortium of partners from AWI (Alfred-Wegener-Institute Helmholtz Centre for Polar and Marine Research), DLR (German Aerospace Center), KIT (Karlsruhe Institute of Technology), and the GEOMAR (Helmholtz Centre for Ocean Research Kiel), we used DTPs \cite{MFI2020}, to develop the ocean observation system without the need of a permanent connection to the hardware. Instead we used emulators of sensors/actuators and virtualized the interfaces. With this approach, we were able to conduct automated SIL testing in a continuous integration/continuous delivery workflow. This method allows teams to share the work easily, and reduce costs for hardware needed solely for development purposes. Furthermore, we were able to develop the software from our home offices during the COVID-19 pandemic. After successful tests in October 2020 in the Baltic Sea \cite{fieldreport}, we were curious how SMEs develop and test embedded software systems. 

\noindent Hence, the goal of this interview study is the following:
\begin{quote}
\emph{\textbf{Goal:} Investigate how small and medium-sized companies in Germany develop smart machines for the Industry 4.0 and which challenges they face in this context.}
\end{quote}

To reach this goal, we elaborated three research questions and answer them using qualitative methods:

\begin{enumerate}[label*=\textbf{RQ\arabic*}, leftmargin=1cm]
    \item How do engineers in SMEs develop embedded software systems for Industry 4.0 applications? \label{rq1}
    \begin{enumerate}[label*=\textbf{.\arabic*},leftmargin=0.5cm]
        \item Which strategies do they prefer? \label{rq1a}
        \item Which obstacles arise? \label{rq1b}
    \end{enumerate}
    \item How do SMEs ensure software quality and customizable products in the Industry 4.0?\label{rq2}
    \item Which level of integration of digital twins have the SMEs implemented? \label{rq3}
\end{enumerate}

We conduct semi-structured interviews and analyze them by performing a thematic analysis as described by \textcite{Braun2006thematicanalysis}.
Industry 4.0  strives for customizable products with a lot size of one~\cite{gilchrist2016industry}, which is only feasible via an appropriate software-based automation.

\section{The Research Design}
To answer the research questions, we reached out to the SMEs via acquaintances, colleagues, or E-Mail. In this work, we present the results for six interviews. The target group were engineers (software, electrical, or mechanical) who develop software for embedded systems in SMEs. We did not focus on the experience level and ignored the gender.

\begin{table}[ht]
\renewcommand{\arraystretch}{1.5}
\centering
\begin{tabular}{p{0.075\textwidth} | p{0.35\textwidth}}
    \multicolumn{1}{c}{\textbf{Code}}      &    \multicolumn{1}{c}{\textbf{Scheme}}  \\ \toprule
     \emph{HMCHM} & Hidden Champion, Mechanical Engineer, CPS, HIL, DM \\
     \emph{MSCHM} & Medium, Software Architect, CPS, HIL, DM \\
     \emph{MSSPM} & Medium, Software Architect, Service provider, SIL, DM \\
     \emph{SSSHT} & Small, Software Engineer, Sensors/Actuators, HIL, DT \\
     \emph{SESHM} & Small, Electrical Engineer, Sensors/Actuators, HIL, DM \\
     \emph{SSPSS} & Small, Software Architect, Service provider, SIL, DS \\
\bottomrule
\end{tabular}
\caption{The coding scheme for the participants.}\label{table:codes}
\end{table}
Due to the Covid19 pandemic, the interviews were conducted in online conferences. We asked all participants whether they allow the recording of the interview for a later transcription or not. If an interviewee did not allow the recording, we took notes and enriched them with additional impressions, afterwards. All interviews took 30 to 45 minutes and the interviewees did not receive the questions beforehand. 

In the following, we reference the interviewees by the codes shown in \Cref{table:codes}. Our participants are working for enterprises that are small (S), medium-sized (M), or are hidden champions (H). By their education they are software engineers/architects (S), mechanical engineers (M), or electrical engineers (E). Since we do not mention the products the companies sell, we distinguish only between cyber physical systems (CPS) (C), sensors/actuators (S), and service provider (P). We categorize the development workflow in SMEs into Hardware-in-the-Loop (H) or Software-in-the-Loop (S). Last, we categorize their usage of the DT paradigm in the subcategories described by \textcite{KritzingerDTBoxes}, namely the digital model (M), digital shadow (S), and digital twin (T). All these characteristics combined provide unique codes for each interviewee.
For instance, HMCHM is a \underline{H}idden champion, \underline{M}echanical engineer, works on \underline{C}yber physical systems, uses \underline{H}IL and digital \underline{M}odels.

\subsection{Interview Questions}
\begin{table*}[ht]
\centering
\begin{tabular}{p{0.1\textwidth} | p{0.7\textwidth}}
\multicolumn{2}{c}{\textbf{General Interview Questions}}  \\ \toprule
    & What is your companies' main product and which role do you have? \\ \midrule
     & Please describe a typical workflow, when you develop/adjust a feature? \\ 
\multicolumn{1}{c|}{\ref{rq1}}          & Do you remember a (complicated) problem you had to solve recently and how did you solve it?\\ 
          & What do you like about your approach and what do you dislike? \\ \midrule
\multicolumn{1}{c|}{\multirow{2}{*}{\ref{rq2}}}          & How do you ensure that new features or bug fixes do not compromise your product? \\ 
          & How do you individualize the software of your product? \\ \midrule
\multicolumn{1}{c|}{\ref{rq3}}         & Are you able to monitor the product from remote? \\ 
\bottomrule
\end{tabular}
\caption{The interview questions mapped to the research questions.}\label{table:questions}
\end{table*}

To answer our research questions, we asked the questions that are listed in \Cref{table:questions}. When we elaborated the questions presented in \Cref{table:questions}, we followed the guide by \textcite{carlaqualitativeresearch}. Notice, that these questions are only a template. Since we conducted semi-structured interviews, they may vary from interview to interview depending on the conversation's flow. For example, we asked ``\emph{You mentioned you are developing XYZ, can you please describe how your development workflow looks like?}'' instead of ``\emph{Please describe a typical workflow, when you develop/adjust a feature?}''. Furthermore, depending on the conversation's flow, follow-up questions could arise that are not listed here, yet are relevant to get to know their development processes. An example for a follow-up question we asked in the interview with \emph{MSSPM} is ``\emph{How is your experience with the simulation you mentioned, does it work as expected or do you have to adjust many things when you switch to the real system?}''

With regards to DTs, we already mentioned that the definition of a DT differs between industrial practice and academia. In the industry, marketing departments advertise IoT-platforms and simulation tools as DTs, despite the fact that the automated data exchange is only in the direction from the physical object to a database and/or the digital model in the simulation. Hence, it is only a digital shadow. A fully implemented DT also reflects the other way round, the automated data exchange from the digital shadow to the physical object. Thus, we asked whether the interviewees can monitor (digital shadow~\cite{KritzingerDTBoxes}) or also control (DT~\cite{KritzingerDTBoxes}) their products.

\subsection{A Digital Twin Prototype of a PiCar-X}
For the exemplary context of automotive software engineering, we illustrated our DTP approach to the interviewees via a SunFounder PiCar-X demonstration. 
To demonstrate some of benefits of using DTs in the development process, we offered all interviewees to present our work in the project ARCHES after each interview. Since our DTP approach is generic, we developed and demonstrated a DTP for low-price hardware around a Raspberry Pi. Therefore, we employed the SunFounder PiCar-X for independent replication of the approach. The software running the PT and its DT are identical. 
The DTP is built upon the Robot Operating System (ROS) and we made the source code available open source on GitHub \cite{abarbiegithub}.

\section{Results and Discussion}
\begin{figure}[ht]
    \centering
    \includegraphics[clip, trim=0cm 11cm 0cm 0cm, width=.475\textwidth]{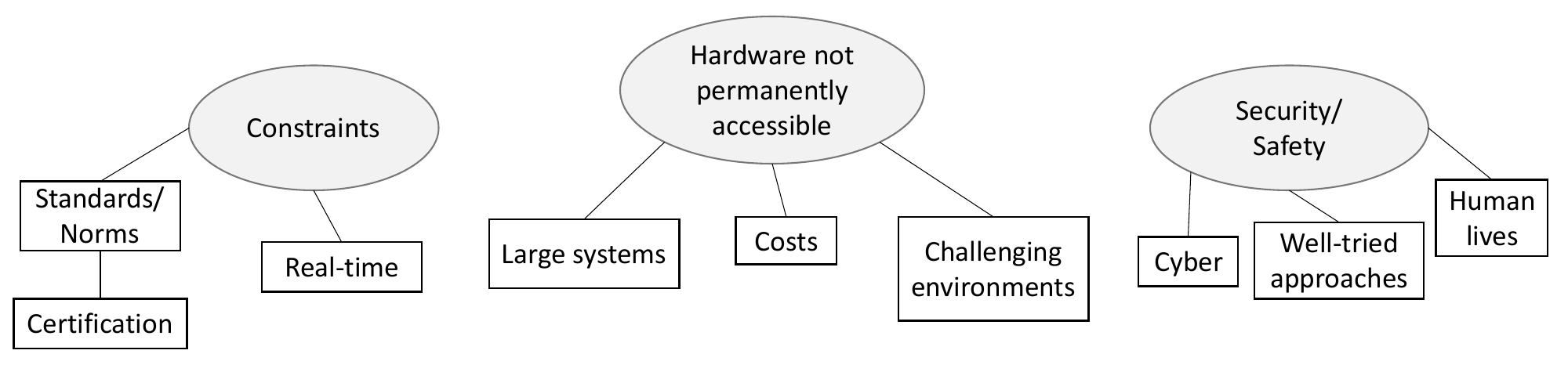}
    \caption{Exstract of the thematic map}
    \label{fig:themes}
\end{figure}

For this study, we chose thematic analysis as the method to analyze our interview data \cite{Braun2006thematicanalysis}.
The central concept of thematic analysis are themes. A theme captures an idea, concept or pattern within the data which seems relevant in relation to the research question.
Due to space restrictions, we cannot present all themes, hence, we focus on the themes presented in Fig.~\ref{fig:themes}.

We started all interviews with an initial question on the enterprise's main product and the role of the interviewee. 
\emph{SSPSS} described the companies' role as service provider as an ``external workbench for companies'' with a focus on SMEs, since ``large companies are starting to set up their own software development departments. SMEs don't have this option.'' This view is underlined by the general demand for software developers in nearly all areas and fields, yet too few new software engineers are trained. Five of the six SMEs we interviewed, were actively looking to recruit software engineers. 

\subsection{RQ1: How do engineers in SMEs develop embedded software systems for Industry 4.0 applications?}\label{subsec:ResultRQ1}
After the warm-up question, we asked all participants to describe their workflow when they develop new software/features. \emph{HMCHM}, \emph{MSCHM}, \emph{SSSHT}, and \emph{SESHM} have access to the hardware for the whole project run-time and they use HIL approaches to develop software. This is different for \emph{MSSPM} and \emph{SSPSS}. 
\emph{MSSPM} described their workflow as follows: ``\emph{Typically, we get the [devices] provided by the project, develop the application, do a few acceptance tests, and then it goes back [to the client]. Then we no longer have a device with us. This is exactly when we start with our simulators. We're not simulating the complete [CPS], but the aspects we need to control the device [...]. That means, we do not only test the individual connection, but we also test the entire system configuration}''. \emph{SSPSS} does have a different reason to switch to a simulation. As a small company they cannot afford spare hardware, hence the hardware prototype the engineers are working on is also used for field tests. Since they are operating in an agricultural environment, the field tests are conducted seasonal from March to October. This is similar to our difficulties developing ocean observation systems, while they operate several months underwater, during which time we do not have any access to the hardware \cite{MFI2020}. This can be reconciled with DTs~\cite{Neto2020}.

We asked \emph{HMCHM}, \emph{MSCHM}, and \emph{SSSHT} as follow-up questions, if they also tried some SIL approaches. All three have hard real-time requirements with regards to their product's software. \emph{HMCHM} and \emph{MSCHM} argued with the size of their systems and the need for hard real-time. \emph{MSCHM} argued with the number of sensors and actuators (``\emph{[several hundred] sensors/actuators are connect to the programmable logic controller}'') and that they are working on simulation tools, but, to date, did not use them for development. Simulating all sensors/actuators and their interfaces in combination with programmable logic controllers (PLCs) and hard read-time requirements is complex. In these cases, the divergence between development environment in a simulation/virtual machine and the execution environment prevents the utilization of DTs not only in the automated test pipeline (see \textcite{Zampetti2022}) but already in the development process. To develop the ocean observation systems with our DTP approach, we emulate interfaces between sensors/actuators and the embedded software system with Docker \cite{MFI2020}\cite{fieldreport}. \textcite{Lyu2021} also demonstrated that a software PLC in a SIL context can be realized with Docker and other tools \cite{Lyu2021}. However, Docker does not reliably meet real-time requirements in industrial contexts. Particularly, in safety critical contexts with hard real-time requirements this is a handicap. While soft real-time requirements can be met with Docker containers \cite{Sollfrank2021}, this cannot be guaranteed for applications requiring hard real-time responses. Overhead and challenges with hard real-time requirements were also a reason why \emph{MSCHM} cannot use Docker to virtualize all components. 

\emph{HMCHM} described the difficulties including assembly lines in a simulation approach, in particular, if not uniformly shaped objects are transported by the assembly lines. He described a HIL approach used by a few colleagues, who connect a PLC to a simulation tool, simulate all the components and then develop the software for the PLC. This approach is quite common today and was also described by \textcite{Lyu2021}. Nevertheless, \emph{HMCHM} did not seem very convinced by this approach. He described cases where a lot of additional effort had to be put into the handling of misshaped objects after the machine was assembled and a strict standardization of the allowed objects is not an option.

Switching from HIL to SIL approaches and convincing engineers and customers of these ideas is not trivial. We asked \emph{MSSPM} about the experience with the simulation. He answered with regard to customers: ``\emph{... you need a standing to simply say we will do it [the simulation]. Because, when I state that I need two man-months for the device connection and another month on top for the simulation, then they asks me `Why do not skip the month}''.
With regards to colleagues, he described the HIL approach at another branch office: ``\emph{The colleagues ..., who are closer to the hardware and also do a lot of projects with their own devices and have a few purchased devices, have a large laboratory and lots of devices. Every time they want to test something, a system has to be plugged together [with hardware] before they can start testing. It took me quite a while before they even understood what we are actually doing here. Because they think, `I will take a device and try it out'} ''. The difficulties in convincing engineers to move from well-tried approaches to new ones, was also described by \emph{HMCHM} in the context of switching interfaces from ProfiBus to ProfiNet. While more senior colleagues prefer the ProfiBus interface, which they use since decades, younger colleagues prefer ProfiNet interfaces. \textcite{Neto2020} listed the convincing aspect as a possible barrier.

\subsection{RQ2: How do SMEs ensure quality and individual products in the Industry 4.0?}
With regard to customizing the software, we found the biggest differences between the interviewees. While \emph{HMCHM} and \emph{SESHM} both ``\emph{unfortunately}'' do not have a modular code basis that could easily be customized or reused, \emph{SSPSS} already has implemented an industrial DevOps approach. The advantage of \emph{SSPSS} is that they rather build an IoT-platform to monitor and operate embedded systems, not the embedded system itself. \emph{SSPSS} described this with: ``\emph{Our IoT-platform is like a big adapter. Sensor can be installed and connected via MQTT, or to OPC-UA, or something else. It is all built into this platform in such a way that I can swap it as components. This means, if a new protocol is invented, then it can also speak to us}''.

Except for \emph{HMCHM}, all interviewees use tools such as SVN or Git for code versioning and automated unit testing. Automated integration testing is only planned by \emph{SSSHT}. Since they do not have access to the physical prototype for long periods, they started to develop a DTP, which then can be tested with an automated SIL approach. Except for \emph{SSSHT} and \emph{SSPSS}, all other interviewees do not have a particular road map to automated integration tests with sensors/actuators in the loop. They all mentioned that industrial standards and certification requirements make it difficult to switch to fully automated approaches. In particular, in domains where malfunctions can endanger human lives, these certification processes are very strict and all parts have to be tested and documented very carefully. This was also mentioned by interviewees in the study by \textcite{vanderValk2021}. \emph{MSCHM} and \emph{SESHM} both explained why they cannot test their systems in an automated way with the requirements for hard real-time responses. However, all participants acknowledged that manually testing takes much time in their development process. To reduce the time for manual test execution, \emph{MSCHM} developed an automated HIL test suite that engineers can start manually on the system to execute all tests on the hardware and monitor them during execution. His colleagues were skeptical first, but saw the benefits and now adopted it in some use cases.

\subsection{RQ3: Which level of integration of DTs have the SMEs implemented?}
\emph{SESHM} uses a digital model only in the development process of the product. \emph{MSCHM} and \emph{MSSPM} both use models but cannot operate their product from remote, due to cyber security concerns. A hacked product could endanger human lives. However, \emph{MSSPM} uses the data of the digital shadow in the development process and \emph{MSCHM} also envisions to use a digital shadow in the development process. With its IoT-Platform \emph{SSPSS} provides a tool for a digital shadow. The only one who plans to operate a full DT is \emph{SSSHT}. 

\section{Threads to Validity}
We identified the following threads to the validity of this qualitative study: (i) only six interviews are presented, (ii) we did not distinguish between the experience level, e.g. junior vs. senior engineers, and (iii) all interviewees are from different industries and hence, the results cannot be mapped to a specific industry.

\section{Conclusion and Future Work}

The result for the research questions \ref{rq1} and \ref{rq3} show that constraints to development processes, security/safety concerns, and the accessibility of hardware are the main themes in these  interviews. This is no surprise, since these points were also found in other studies, which we discussed in Section~\ref{sec:relatedwork}. Limited access to the hardware leads to SIL development approaches based on simulations/emulators. Yet, this is not in all use cases possible. There is a lack of proper tooling in applications that have hard real-time requirements. Not until these tools exists, DTs can be fully integrated in the development process of embedded software systems with (hard) real-time requirements. All interviewees see the potential of SIL approaches and DTs with regard to quality and customization (\ref{rq2}), although it will take some effort to convince and establish SIL development with DTs in the different SMEs. 

One reason it will take some effort to convince engineers, could be the conservative nature of the embedded community, particularly in SMEs. We observe that while large companies often already have an established technology stack including Git and automated CI/CD pipelines used by software engineers outside the embedded software field, SMEs often have to built up some software engineering competence from scratch. Although software engineering is an increasingly important for all engineering domains, the education outside of computer science courses still lacks behind the State-of-the-Art. With a better knowledge of software engineering methods, tooling, and best practices, some of the hurdles could easily be taken. As \emph{MSCHM} described, engineers sometimes just have to experience the benefits of new methods. Introducing new tools is easier in smaller groups with less stakeholders and decision makers.

All in all, digital twins have the potential for cost-effective software development and maintenance strategies \cite{ArchitecturalConcerns}. Furthermore, automated processes help to increase the software quality of embedded software and also have the potential to assist engineers in robust and faster certification processes. All these benefits are in favor of the development processes in SMEs, although some education is needed. Nevertheless, to fully unfold the advantages of digital twins, more research is required for more advanced tools and methods with real time capabilities. This does not only challenge the traditional software engineering community, but in particular the embedded software community that needs to adapt to a fast changing software development environment. How to master the transition from a conservative embedded software engineering style to fully integrated digital twins got already a lot of attention by academia and practitioners, yet there is still plenty of research to do. Adding more software engineering methods to the education of all engineering fields is a trivial postulation. More challenging will be the education of engineers that are already in the middle or near the end of their professional life and still need to adapt to the growing changes. Due to the demographic change in many 
industrial countries this group is larger than the group of engineers finishing university and starting their professional careers. How to achieve this, has to be researched extensively. 

To get a deeper insight into the development workflow of SMEs, we will conduct more interviews to extend and deepen our analysis.

\section{Acknowledgments}
This project was supported through the Helmholtz Enterprise Field Study Fellowship within the project `Paladin Embedded Software Systems'.

%
%
%
\printbibliography
\end{document}